\documentclass[%
reprint,
amsmath,amssymb,
aps,]{revtex4-1}
\usepackage{graphicx}
\usepackage{dcolumn}
\usepackage{bm}
\begin{document}
\title{Solar Wind Bremsstrahlung off DM in our Solar System}
\author{Gilles Couture}
\affiliation{D\'epartement des Sciences de la Terre et de l'Atmosph\`ere\\
Universit\'e du Qu\'ebec \`a Montr\'eal\\
201 Ave. du Pr\'esident Kennedy, Montr\'eal, QC\\
Canada H2X 3Y7}
\date{\today}
\begin{abstract}
We consider the possibility that the solar wind emits photons via 
{\it bremsstrahlung} when colliding with dark matter (DM) particles 
within the solar system. To this effect, we calculate the bremsstrahlung 
spectrum a proton would emit when colliding with a neutral spin-1/2 particle 
through the exchange of a scalar neutral particle. We assume a speed of 600 km/sec 
for the solar wind and assume that the speed of the dark matter halo is due to the 
motion of the sun through our galaxy, which we take as 300 km/sec. We assume a DM density 
of $0.3~GeV/cm^3$ and a solar wind composed primarily of protons with a total rate of 
ejection mass set at $10^9kg/sec$. We use a Monte Carlo technique to let this interaction 
take place within the solar system and calculate the photon rate an 
observer would detect on Earth or at the edge of the solar system as a function of 
photon energy. We find that the rates are in general very small but could be observable in some 
scenarios at wavelengths in the $mm$ or $cm$ range.   
\end{abstract}
\pacs{to come}

\maketitle

\noindent{\bf {Introduction}\hfil\hfil}
\hfil\hfil\break\noindent
About one year ago, Xenon1T reported a clean signal above known backgrounds [1]. This measurement 
triggered a flurry of activity aiming at explaining this signal in the context of 
Dark Matter (DM)[2-5] or others [6,7].  Xenon1T is one of several {\it direct detection} 
cryogenic detectors [8-13] where the DM particle interacts directly with the nucleus of the 
detector and one records its recoil [14,15,16]. 
There are also direct detectors where the energy deposited by the DM 
particles causes the explosion of a superheated droplet embedded in a gel and one simply records the 
sound wave [17-20]. The direct detection approach suffers from a mass 
threshold: when the mass of the DM particle becomes small, it becomes very difficult for the 
DM particle to deposit enough energy in the nucleus to produce a detectable signal. 
In order to overcome 
this low mass threshold, new mechanisms based on the Migdal effect use the effect of the 
recoil of the nucleus on the electronic cloud [21-30]. 
Another avenue is to explore the possibility that the 
DM particles be upscattered by more energetic particles such as cosmic rays.[31,32]

The indirect detection approach relies on the gravitational pull of massive bodies that could 
lead to the 
accumulation of DM particles in their core, be they the Earth, the sun or the galaxy. 
This increase in 
density will lead to an enhanced probability of annihilation DM-particle-antiparticle which 
would then 
produce standard particles that could be detected on Earth.[33] 
The presence of DM inside the sun would have an effet on its
behaviour and some interesting limits on axion-photon couplings have been obtained from a global 
fit to solar data, including helioseismology [34]. Allowing both the accumulation of DM particles 
inside the sun and their interaction with electrons would lead to DM upscattering off the electrons 
and possibly interesting rates on Earth of DM energetic enough to produce a signal in the direct 
detection experiments [35]. A similar accumulation could take place within exoplanets and would 
affect their evolution [36]

The presence of DM in the early universe will also have an effect on its early evolution [37]
and could produce a distortion in the cosmic microwave background at recombination [38] as well as
affecting the evolution of galaxies and their satellites later on [39].

Recently, an Earth-scale detector called GNOME [40] reported a first null result from a relatively 
short operation time aiming at measuring a topological defect due to an axion-like particle field
propagating through space [41]. The signal from such a field as the Earth crosses it would be a 
slight perturbation in high precision/sensitivity instruments such as atomic clocks occuring at 
slightly different instants.

DM particles could also be produced at high energy colliders such as the LHC [42,43]

In this paper, we calculate the bremsstrahlung photon flux one would receive 
from the collision between a proton from the solar wind and a DM particle in the vicinity 
of the solar system. We consider space itself as a collider whose collision angle and 
luminosity will vary throughout the solar system. We neglect cosmic magnetic fields and 
assume that the solar wind travels in straight line at a speed of 600 km/sec 
($\beta_{_M} = 0.002$). We consider that protons constitute most of the solar wind and 
the ejection mass rate from the sun is set at $10^9~kg/sec~=~{\cal{R}}_{sw}$. 
We assume that the dark matter has a density of $0.3~GeV/cm^3$ [44,45,46] and that it 
is at rest in our part of the galaxy; its relative motion is due to the motion of the sun, 
which we take as 300 km/sec ($\beta_{_{DM}} = 0.001$)

In the following sections, we describe some technical points before presenting our 
results and conclusions. 
\hfil\hfil\break
\noindent{\bf {Procedure}\hfil\hfil}
\hfil\hfil\break\noindent
The process we are interested in is
\begin{equation}
p(p) + q(q) \to p(\bar p) + q(\bar q) + \gamma(k)
\end{equation}
\noindent
where $p$ is the proton, $q$ is the DM particle and $\gamma$ is the photon. 
As it is a three-body final sate, we will use a Monte Carlo technique to calculate this 
spectrum.

{\it Bremsstrahlung} has been studied by Bethe and Heitler in the context of a charged 
particle colliding with a much more massive particle; for example, a proton impinging 
on a heavy nucleus [47,48] This process is still of interest as higher corrections are 
added to the original work [49-54] Processes where a hard photon is produced through the 
interaction of a particle with a DM particle have been studied recently [55,56], but the bremsstrahlung 
we study here is a little different in that it is due to the exchange of a light neutral particle. 
When a proton collides with an atom, it does not feel an important positive charge  
until it is well inside the electronic cloud and feels the full nuclear electric charge when 
it is inside the innermost electronic shell. This takes place at a distance that is less than 
0.053 nm since the average value of $r$ is ${a_0\over 2Z}(3 n^2-l(l+1))$. 
With DM, it does not see the object until it is within the range of 
the messenger scalar, which is about $2\times 10^{-15}~m$ for a messenger of 0.1 GeV.

In order to calculate the photon flux, we will proceed as follows:
\hfil\hfil\break\noindent
- using a Monte Carlo technique based on the matrix element of the process at hand, 
we build the distribution $d^2\sigma/dE_\gamma~dcos\theta_\gamma$ 
for several collision angles between the proton and the DM particle
\hfil\hfil\break\noindent
- we pick an observation point with an observation cone of a given opening angle 
and orientation: we will consider 2 observation points and 4 orientations.
\hfil\hfil\break\noindent
- using a Monte Carlo technique, we scan the observation cone in a random and uniform way. 
Each observation point within this cone defines completely all angles and distances of the 
collision.
\hfil\hfil\break\noindent 
- since we have calculated $d^2\sigma/dE_\gamma~dcos\theta_\gamma$ 
for that given colliding angle and outgoing photon angle, we select $\sigma$ at random 
within the distribution and read the corresponding $E_\gamma$ for this event.
\hfil\hfil\break\noindent
- multiplying this partial cross-section by the luminosity, ${\cal L}$ at that position 
and different geometrical parameters that take into account the fact that the photon can 
be emitted anywhere on a cone of angle $\theta_\gamma$ and will not necessarily reach the Earth,  
we obtain $\Phi_\gamma$ and $E_\gamma$ for this event.
\hfil\hfil\break\noindent
- averaging the distribution produced within the cone and multiplying by the number of independant 
colliders in the cone, given by  $V_{cone}/V_{accelerator}$ will give the desired flux.
\hfil\hfil\break\noindent

We set our axes such that the sun 
is at the origin, the Earth is along the positive y-axis and its rotation around the 
sun is from the positive-x to the positive-y axes.

\hfil\hfil\break
\noindent{\bf {Matrix Element}\hfil\hfil}
\hfil\hfil\break\noindent
We consider a relatively simple, generic model where a proton scatters 
off a neutral spin-1/2 DM particle through the exchange of a scalar 
neutral DM particle. We set the mass of the proton at 1 GeV. 
The DM masses are free and the couplings are also free:
$p-p-DM_{exchange}=\lambda$ and
$DM_{collide}-DM_{collide}-DM_{exchange} = \Lambda$, where $DM_{collide}$ is the DM particle that
collides with the proton and $DM_{exchange}$ is the DM particle exchanged between the proton
and the $DM_{collide}$ particle.
Since we do not take into account the spin of the proton, the only angles that can be defined 
in the process are with respect to $\vec\beta_{_M}$: ~$\theta_{_{DM}}$, the angle between the 
incoming proton and the DM particle, and $\theta_\gamma$, the angle of the outgoing photon with 
respect to the incoming proton are the relevant angles of the process since we do not observe 
nor the proton nor the DM particle after the collision.

As the speeds of the colliding particles are small, we have a non relativistic problem where 
the maximum energy of the photon is
\begin{equation}
E_\gamma^{max} = 
{(1/2)(M_{_M}~M_{_{DM}})(\vec\beta_{_M} -\vec\beta_{_{DM}})^2\over M_{_M}+M_{_{DM}}}
\end{equation}
\noindent
where $M$ as a subscript stands for $matter$ or $proton$ in ou case.

We are mostly interested in the low energy part of the photon spectrum: 
the eV and sub-eV scale. As is well known, [57,58,59]
bremsstrahlung processes tend to diverge if we let the energy of the photon 
go to 0, the divergencies being cancelled by the higher order corrections. 
In our case, we set the minimal energy of the photon at $10^{-10}$ GeV. 
The null mass of the photon is potentially a numerical problem; in 
order to avoid numerically negative photon masses and instabilities, we give a mass to the 
photon ($k^2 = m_\gamma^2$) and include the extra terms in the matrix element that come with this 
mass: the summation over polarisation states is ($g^{\mu\nu}-k^\mu k^\nu/m_\gamma^2$).  
We have verified that masses between $10^{-13}~GeV$ and $10^{-16}~GeV$ 
give identical results but a mass of $10^{-11}~GeV$ does not; we worked with 
$10^{-14}~GeV$

At this point, when considereing a head-on collision process and varying the incoming 
energies of the colliding particles, we observe the usual behaviour of bremsstrahlung photons as having a 
slightly enhanced emission probability at right angle at low energy and peaking more 
and more in the forward direction at higher energies.

\hfil\hfil\break
\noindent{\bf {Monte Carlo}\hfil\hfil}
\hfil\hfil\break\noindent
Clearly, in the process that we want to study, 
the solar wind will not be in a head on collision with the DM. We then 
modifiy the boost factor of our center of mass system (CM) to take into account 
the angle of incidence of the DM. [48,60,61] In every collision, the motion of the solar wind particle
defines our positive $z^\prime$-axis. The angle between the matter particle and the DM particle, 
$\theta_{_{DM}}$ is defined such that a head-on collision corresponds to $\theta_{_{DM}} = 0$ 
and collinear collision corresponds to  $\theta_{_{DM}} = 180$. 
We then consider a boost factor at a given angle $\theta_{_{DM}}$:
\begin{equation}
\vec\beta_{CM} = (-q_{x^\prime}, -q_{y^\prime}, p_{z^\prime}-q_{z^\prime})/(p^0+q^0)
\end{equation}
\noindent
and
\begin{equation}
\chi_{CM}= (1/2)~log((1+\beta_{CM})/(1-\beta_{CM}))
\end{equation}
\noindent
where $\chi$ is the rapidity factor of the process. 
The Mandelstam variable $s\equiv(p+q)^2$ becomes 
\begin{equation}
s= M_{_M}^2 + M_{_{DM}}^2 + 2 M_{_M}M_{_{DM}}\gamma_{_M}\gamma_{_{DM}}
(1+\beta_{_M}\beta_{_{DM}}cos(\theta_{_{DM}}))
\end{equation}
\noindent
where $\gamma_{_M}$ and $\gamma_{_{DM}}$ are the usual relativistic parameters.

Taking these into account, we use a Monte Carlo technique to build the different distributions.
The distributions are 
$ d\sigma/dcos(\theta_\gamma), 
d\sigma/dE_\gamma,~{\rm and}~d^2\sigma/dcos(\theta_\gamma)~dE_\gamma$; where $E_\gamma$ is the energy 
of the emitted photon and $\theta_\gamma$ is the photon angle with respect to the positive 
$z^\prime$-axis, {\it ie} the incoming matter particle. The energy distribution 
converges very smoothly, but the angular distribution, due to the three-body final state,  
is slower to converge and required $2\times 10^9$ events to reach 
a relatively clean distribution. There was still a fairly large spread in the 
amplitude of the angular distributions and we used a fifth-order smoothing algorithm  
[62]:
\begin{equation}
y_i = {1\over 35}(-3 y_{i-2} + 12 y_{i-1} + 17 y_i + 12 y_{i+1} -3 y_{i+2})
\end{equation}
\noindent 
After 150 iterations, the spread is reduced by about a factor 5 and keeps the general features 
of the original distribution as can be seen on figure 1 where the narrow band is the final distribution.

The doubly differentiated distribution 
$d^2\sigma/dcos(\theta_\gamma)~dE_\gamma$ is the most important in what follows. 
In order to study the interaction of the solar wind with the DM halo, we 
built this distribution for collision angles ($\theta_{_{DM}}$) from 0 to 180 degrees by increments 
of 10 degrees; we have then 19 distributions.

\hfil\hfil\break
\noindent{\bf {Scanning the solar system}\hfil\hfil}
\hfil\hfil\break\noindent
Once we have these distributions, we let the solar wind scatter off the DM in the solar 
system and beyond. We first pick an observation point from which originates an observation 
cone of a given opening angle ($\delta_C$), length and orientation; all photons observed must 
be produced within this cone. Once we have picked a collision point at random within the cone, 
all angles and distances are well defined: $\theta_{_{DM}}$, $\theta_\gamma$, $\vec R$ (the vector from 
the observation point to the collision point) and $\vec D$ (the vector from the sun to the 
collision point).

\noindent
We build the photon flux ($\Phi$ in $1/m^2~sec$) as
\begin{equation}
{d\Phi\over dE_\gamma} = {1\over N}\bigg\langle{d\Phi\over dE_\gamma}\bigg\rangle~\left({V_{cone}\over V_{collider}}\right)
\end{equation}
\noindent 
where
\begin{equation}
\bigg\langle{d\Phi\over dE_\gamma}\bigg\rangle = \bigg(\sum_{i=1}^N\left({d^2\sigma\over 
dcos(\theta_\gamma)dE_\gamma}\right)_i\cdot G_i
\bigg)\bigg\vert_{{cos(\theta_\gamma)_i =\atop cos(\theta_\gamma^{Earth})_i}}\bigg\vert_{\hat r_i\in cone}
\end{equation}
\noindent
and 
\begin{equation}
G_i = {(\Delta cos(\theta_\gamma))_i~{\cal L}_i\over 2~\pi~(R~sin(\theta_\gamma))_i~(R~\Delta\theta_\gamma)_i)}
\end{equation}
\noindent
${\cal L}_i$ is the luminosity at the colliding point, ${\hat r}_i$ is a unit vector from the 
observation point to the colliding point, $\Delta cos(\theta_\gamma)$ is the width of the bin previously 
defined, and $\Delta\theta_\gamma$ is the angular width of the band that we consider at the Earth. $G_i$ takes into account the fact that the photon is emitted on a cone of angle $(\theta_\gamma)_i$ and will not necessarily reach the Earth. Once the cone 
has been sampled and averaged, we multiply by the number of independant accelerators that we have in our cone.

The sun is at the origin of our coordinate system and our observation point is along the positive $y$ axis; 
this is $\vec R_o$.  
We define the elevation angle of the observation apparatus as $\theta$ above the ecliptic 
plane and $\phi$ from the positive $y$-axis towards the positive $x$-axis. The aperture of 
the observing apparatus defines the opening angle of the observing cone; $\delta_C$ such that the total 
angle of the cone is 2$\delta_C$. We will use $\delta_C= 40^\circ$.

In order to sample the cone in a uniform and random manner, we embed the cone in a cube 
($x'', y'', z''$) tailored to the cone and produce random numbers within each axis of 
the cube. We keep the events that fall within the cone and do not take into account those that lay 
outside of our cone. Due to the $1/D^2$ behaviour of the solar flux that will 
appear later in the luminosity and the $1/2\pi~R~sin\theta_\gamma~R~\Delta\theta_\gamma$ behaviour of the area 
of the ring where the photon can be emitted, one expects that the contribution of very far away colliders 
will decrease and therefore, when building our distributions, we should reach a plateau, a maximum photon 
flux once the length of our cone of integration reaches a certain value. Numerically, we should even 
observe a decrease in the amplitude of the distributions if we increase the length of the cone 
since integration up to very far distances would take a very large 
number of events in order to scan properly the regions that contribute most. We observe 
this behaviour in our distributions. We should also observe an increase in the number 
of photons when we increase the opening angle of the observational cone; this relation is not so 
straightforward though as the emission of the bremmstrahlung photon is not uniform in the collision 
process and some regions of the observation cone might be better at producing photons than others.  
We also observe this behaviour in our distributions.

In order to scan the cone, we proceed as follows: our observation point is at the origin of our  
$x''-y''-z''$ frame where we generate 
coordinates at random. This frame has the same orientation as our observation cone and its 
$x''$-axis corresponds to our $x$-axis (we set $\phi=0$). 
The cone is simply rotated by a certain angle around its 
$x''$-axis. Once the $(x'',y'',z'')$ coordinates are generated at 
random, we have $\vec R$, the vector from the observation point to the colliding point.   
It is then straithgtforward to define the vectors that define the position of the 
observation point and the collision point. We consider four such rotations: 
\hfil\hfil\break\noindent
A- 45 degrees: we look 45 degrees above the ecliptic with our back to the the sun
\hfil\hfil\break\noindent
B- 90 degrees: we look 90 degrees above the ecliptic plane
\hfil\hfil\break\noindent
C- 135 degrees: we look 45 degrees above the ecliptic plane and partly toward the sun
\hfil\hfil\break\noindent
D- 180 degrees: we look directly at the sun but we exclude a cone of 2 degrees full angle in order 
to exclude the sun itself.

The relevant vectors are then:
$$
\vec{R_o} = (0,R_o,0)\phantom{sssss}\vec{D} =\vec{R_o}+\vec R
$$
$$
\vec V_{_{DM}}= \beta_{_{DM}}(0,-1,0)\phantom{sssss}
\vec V_{_M}= \beta_{_M}(D_x,D_y,D_z)/ \vert\vec D\vert\hfil\hfil
$$

In this work, we assume that the DM is incoming along the $-y$ axis; the motion of the sun is 
in the direction of the ecliptic plane towards the Earth. 
The angles $\theta_{_{DM}}$ and $\theta_\gamma$ are obtained from the scalar products
$$
cos(\psi) = {{\vec D}\cdot{\vec V_{_{DM}}}\over D~V_{_{DM}}}~{\rm and}~
cos(\eta) = {{\vec D}\cdot{\vec {R}}\over D~R}
$$  
From these definitions we have $\theta_{_{DM}} = \pi-\psi$ and $\theta_\gamma = \pi-\eta$.

\hfil\hfil\break
\noindent{\bf {Probability for the outgoing photon to be ejected with a given energy}\hfil\hfil}
\hfil\hfil\break\noindent
Once these angles are determined we go back to the distributions we 
built previously and pick the one that corresponds to the correct value of $\theta_{_{DM}}$;  
{\it ie} the $d^2\sigma/dcos(\theta_\gamma)dE_\gamma$ that corresponds to our value of 
$\theta_{_{DM}}$. Within this distribution, we pick the 
specific row that corresponds to the correct value of $\theta_\gamma$. 
This row represents the distribution $d\sigma/dE_\gamma$ for $\theta_{_{DM}}$ and $\theta_\gamma$ fixed 
by the position of the collision point. 
Within this distribution, we consider the highest possible value of the differential 
cross-section and pick a number at random between 0 and 1 to mutiply it with. 
Since the cross-section (differential or total) represents the probability that a given process 
takes place, multiplying this maximum value by a random number between 0 and 1 will give the probability 
that the wanted process take place. We then read the corresponding value of the photon-energy. 
We now have the cross-section (or probability) that a photon be produced at a given angle with a 
given energy when a proton of a given energy collides with a DM particle of a given energy at a 
given incoming angle.

\hfil\hfil\break\noindent
\noindent{\bf {Luminosity}\hfil\hfil}
\hfil\hfil\break\noindent
The luminosity ($1/m^2~sec$) depends on the densities and velocities of the colliding particles 
and also on the collidng angle. [63,64]
We assume a uniform density of the dark matter cloud within our region 
of the galaxy and set it at $0.3~GeV/cm^3$. 
Regarding the solar wind, we assume that the sun emits a certain amount of material in space 
every second: we set it at $10^9$ kg/sec and assume that it is mostly protons. We also include the usual 
$1/D^2$ behaviour of the solar flux, which leads to a decreasing density of the solar wind but we 
assume that this density is constant over the 600 km that our solar wind travels in 
one second. Essentially, we consider a beam of uniform density over 600 km in length and 1 meter 
in cross-section. Since the interaction rate is rather small, we do not take into account the 
depletion of the solar wind flux as it travels through space.  
In these conditions, the required kinematical factor is given by
\begin{equation}
K=\sqrt{(\vec\beta_{_M}-\vec\beta_{_{DM}})^2 - (\vec\beta_{_M}\times\beta_{_{DM}})^2}
\end{equation}
\noindent
so that the luminosity is given by
\begin{equation}
{\cal L} = \rho_{_M}~\rho_{_{DM}}~L_x~L_y~v_{_{M}}~v_{_{DM}}~\Delta t~*~ K
\end{equation}
where $L_x$ and $L_y$ are the tranverse sections of the {\it beam}, which we take as 1 m, and 
$\Delta t$ is the time scale of the collision, which we take as 1 second. $\rho_{_M}$ is the 
density of matter (protons) at the collision point, given by 
$$
\rho_{_M} = {{\cal{R}}_{sw}/m_{proton}\over 4 \pi D^2~v_{_M}}
$$
\noindent
and $\rho_{_{DM}}$ is the density of dark matter particles at the collision point in $1/m^3$.


In order to get the final spectrum from this observation cone, we divide the distribution that we just built 
by the number of events used (which gives us the average spectrum) and then multiply by the number of independant 
accelerators within this cone: $V_{cone}/V_{collider}$. We take $V_{collider} =(600000 + 300000)m^3$. 
This procedure is justified since the distances travelled by our particles in 1 second are very much smaller than 
the distances at hand: we would need to consider $\sim 10^{28}$ collision points for our colliders to begin to 
overlap within the cone. 
This procedure is also symmetric both from the point of view of the matter and from the point
of view of the DM particle. We neglect the effect encountered when $\theta_{_{DM}} = 180$, in 
which case the two volumes would overlap.

\hfil\hfil\break
\noindent{\bf {Free parameters}\hfil\hfil}
\hfil\hfil\break\noindent
There are 5 free elements in this scenario: 2 DM masses, 2 DM-couplings and the density of
DM particles in the galaxy. Clearly, the couplings and the density of the DM particles in the
solar system simply factor out as $\rho_{_{DM}}$ and $\lambda^2\Lambda^2$. The mass of the DM
particle exchanged in the process also factors out as $1/m^4$ since we have a t-channel propcess
and the photon energies involved are much smaller than the masses. Regarding the mass of the
incoming DM particle, it does not factor out and we find that the maximum bremsstrahlung occurs
when its mass is about twice that of the proton so that the proton and the colliding DM particle 
have about the same momentum. The angle of the observation cone also factors
out, but one has to be careful because the production volume might depend on the orientation when
this angle becomes small.
Therefore, our results can be scaled up or down by multiplying them by the following expression:
$$
{(\lambda^2~\Lambda^2)\cdot(\rho_{_{_{DM}}}/\rho_{_{_{DM}}}^0)\cdot ({\cal{R}}_{sw}/{\cal{R}}_{sw}^0)\cdot
(tan(\delta_C)/tan(\delta_C^0))^2
\over{(M_{exchange}/0.1 GeV)^4}}
$$
\noindent
where ${\cal{R}}_{sw}^0 = 10^9kg/sec, \rho_{_{_{DM}}}^0 = 0.3 GeV/cm^3, \delta_c^0= 40^{\circ}$

\hfil\hfil\break
\noindent{\bf {Results}\hfil\hfil}
\hfil\hfil\break\noindent
We consider two observation points: one at 1 au (the Earth) and another at the edge of the solar 
system, at 50 au (Note that when looking directly at the sun from 50 au, reducing the exclusion cone from 
1 degree, as it was at 1 au to 1/50 degree has very little effect). 
We also consider two scenarios:  in the first one (scenario A), the colliding DM 
particle has a mass of 2 GeV and the exchange particle has a mass 0.1 GeV. 
Such scenarios have been
considered as {\it secluded WIMP dark matter} and some models allow excited state. [65,66,67].
In the second scenario (scenario B), the colliding particle and the exchange particle have the same mass.  
In scenario A, we use the mass of the 
colliding particle to calculate the density of DM particles ($\rho_{_{DM}}$).
We also considered several lengths and opening angles 
of the observational cone and verified that our distributions behaved as expected.

The result is figures 2, 3, 4, and 5, which 
represent ${d\Phi_{\gamma}\over dE_{\gamma}}~vs~E_\gamma$ where 
$d\Phi_{\gamma}\over dE_{\gamma}$ is in $1/m^2~sec~eV$ and $E_\gamma$ is in $eV$; recall that the visible 
spectrum is from 1.6 eV to 3.2 eV. One notes that:
\hfil\hfil\break\noindent
- all curves exhibit a straight line at low photon energy but show some noise at higher energy; the 
higher energy part of the spectrum converges more slowly. One obtains such straight lines from the classical 
Bethe-Heitler spectrum [50] if plotted on log-log axes.
\hfil\hfil\break\noindent
- the rate at 45 degrees is substantially smaller than the rates at 90, 135 and 180 degrees
\hfil\hfil\break\noindent
- the rates at 90, 135 and 180 degrees are very similar 
\hfil\hfil\break\noindent
- all slopes at 1 au are very similar, at about -3.2 $1/m^2~sec~eV^2$, for both scenarios
\hfil\hfil\break\noindent
- the slopes at a distance of 50 au are also very similar to those at 1 au
\hfil\hfil\break\noindent
- the rate at 50 au when looking at the sun is similar to the rate at 1 au and 45 degrees with a behaviour 
similar to that of 1 ua when the observation angle changes.

The straight lines indicate that 
\begin{equation}
d\Phi(E_\gamma)/dE_\gamma = \Upsilon_0~(E_\gamma/E_\gamma^0)^{-\alpha} \to 
\end{equation}
\begin{equation}
\Phi\big\vert^{E_\gamma^{max}}_{E_\gamma^{min}} = {\Upsilon_0(E_\gamma^0)^\alpha\over \alpha - 1}
\bigg\{{1\over ({E_\gamma^{min}})^{\alpha-1}} - 
{1\over ({E_\gamma^{max}})^{\alpha-1}}\bigg\}
\end{equation}
\noindent
where $\Upsilon_0$ ($1/m^2~sec~eV$) and $E_\gamma^0$ is some reference point.

From figures 2, 3, 4, and 5, we can extrapolate safely to 0.01 eV and with caution somewhat below. 
Using these different parameters, we obtain Table I, where we give the photon flux ($1/m^2~sec$) for 
different scenarios and photon energy bands. As we have already covered the whole observation cone, taking 
into account its opening angle of $40^\circ$, we could say that our units are $1/m^2~sec~(0.47 \pi sr)$. 
One has to be careful in scaling because opening or closing  the observation cone at different observation 
angles might not give the same result as the productive zones may vary in size at different observation 
angles. From Table I, we can see that the rates are very small over the whole range with the parameters used 
here but could become interesting at wavelengths of $mm$ or $cm$ (with energies in the $meV$ range) in 
scenario B.
Clearly, if we reduce the mass of the exchange particle by a factor of $10$, we gain a factor $10^4$ in 
the rates and these become interesting in the microns range, as long as we remain in scenario B. 
In a model where several DM particles can coexist with very different masses and interact 
with each other and with the proton, the scenario that we consider here would be the dominant one: a heavier 
exchange particle would reduce the t-channel amplitude and a lighter colliding particle would reduce the 
bremsstrahlung. Of course, a complete calculation with all diagrams involved would be necessary.

Our results indicate also that the excess luminosity observed recently by the New Horizon probe [68] cannot 
come from this process, at least not with reasonable parameters. 
One would need far higher energy protons (or other
charged particles) or far higher density in order to produce the excess in the visible range; 
likely, such particles would have been observed in other ways.

Reversing the situation, a scenario where the heavier DM particle could couple to lighter dark particles 
would lead 
to their production through bremmstrahlung off the DM particle as it scatters off the solar wind, thereby 
increasing their presence in our solar system. This effect would be larger in close proximity of the sun. 
Similarly, in a model similar to the one considered here, a proton could emit via bremsstrahlung a light 
dark matter particle.[69]

There is also the interesting possibility that an electron could interact with the DM; such couplings have been 
considered before in the context of the 0.511 MeV emission from the center of the galaxy [70,71] and the 
possibility of DM upscattering in the sun [35] 
This process should lead to results similar to what we have here and opens up the possibility of sensitivity 
to very small masses in the DM spectrum since the 
bremsstrahlung seems to be maximum when the colliding particles have about the same momentum and the 
electrons in the solar wind have a speed similar to that of the sun in the galaxy. 
A proposed future MeV gamma-ray telescope is expected to probe the MeV or sub-MeV DM mass range.[72]

We have neglected cosmic magnetic fields, which would bend the solar wind and modify the colliding angles 
within the observation cone. Evaluating the effects of these fields on the spectrum would be interesting, 
but it is unlikely this could make the signal observable in the visible, for example. We have also 
neglected the effect of the sun on the density of the DM. It is likely that the presence of the sun would 
increase 
the density of the DM particles in its vicinity and as most of the spectrum is produced relatively close to the 
sun, this would tend to increase the counting rate observed, but likely not enough to gain several orders of 
magnitude. When considering the spectrum at longer wavelengths however (in the $mm$ or $cm$ range) and  
assuming that the spectrum keeps a slope similar to what we have calculated here, then these 
effects could make a difference in the observable spectrum. 
We have assumed that the sun moves towards the Earth (or the observation points) in its motion around the 
galaxy. Taking the real motion of the sun into account could have a small effect on our results and bring 
some priodicity to our signal. A more precise calculation (finer angular and energy bins and finer sampling 
mesh) is necessary in order to assess the importance of these effect at very low photon energy as well as a 
precise modeling of the cosmos in this regime. [73]

\hfil\hfil\break
\noindent{\bf {Conclusions}\hfil\hfil}
\hfil\hfil\break\noindent
We have considered the scattering of the solar wind off DM particles that might be populating 
our solar system. We allowed the scattered proton to emit a photon through bremmstrahlung and 
calculated the spectrum that one would observe either at 1 au or at 50 au from the sun. We have assumed a 
uniform DM density in our region of the galaxy and neglected the effects of cosmic magnetic fields on the
solar wind. We have found that the rates are very small in general and could not explain the excess luminosity 
observerd recently by the New Horizon probe with reasonable parameters. 
Extrapolating our results down to 0.01 eV is  
reasonable and indicates that the rates are still very small with the parameters we used. 
Reducing the mass of the exchange 
particle to 0.01 GeV could produce a few hundred counts/sec in the 0.01-0.05 eV window (25-125 microns). 
Extrapolating our results to lower energy photons, there could be a measurable 
photon rate in the $mm$ or $cm$ range where one could expect a few hundred to a few 
thousands $photons/m^2~sec$ in the scenario where the exchange DM particle has a mass of 0.1 GeV.  
A more precise numerical calculation 
would be required to confirm this behaviour of the photon flux down to very low photon energies.

The scenario that we have considered and appears to be the most promising is that of a DM particle 
whose mass is about $2~m_{proton}$ 
interacting with a proton via the exchange of a much lighter DM. This opens up the possibility of 
bremsstrahlung of the lighter DM particle off the heavier DM particle as the latter scatters off a proton; 
thereby increasing the abundance of the lighter DM particle.

Considering bremsstrahlung emission from the scattering of electrons off DM particles also opens up the 
possibility of exploring small mass regime in the DM sector since the bremsstrahlung seems to be maximum when 
the momentum of the colliding particles are about the same and the solar wind has a speed similar to that of 
the sun in the galaxy.

\hfil\hfil\break
\noindent{\bf {Acknowledgements}\hfil\hfil}
\hfil\hfil\break\noindent
I want to thank my colleagues Ch\'erif Hamzaoui and Manuel Toharia for stimulating discussions and my colleagues 
from the Atlas Collaboration at the Physics Department at Universit\'e de Montr\'eal for the use of 
their computers.

\hfil\hfil\break
\noindent{\bf {References}\hfil\hfil}
\hfil\hfil\break\noindent
\noindent
{\it  1-}~E.Aprile {\it et als},Phys.Rev.{\bf D102}(2020)072004;
\hfil\hfil\break\noindent
hep-ex:2006.09721
\hfil\hfil\break\noindent
{\it  2-}~H.An, M.Pospelov, J.Pradler, A.Ritz, Phys.Rev.
\hfil\hfil\break\noindent
{\bf D102}(2020)115022;hep-ph:2006.13929
\hfil\hfil\break\noindent
and references therein.
\hfil\hfil\break
{\it  3-}~S.Vagnozzi,L.Visinelli,P.Brax,A.-C.Davis,J.Sakstein, 
\hfil\hfil\break\noindent
hep-ph:2103.15834 and references therein
\hfil\hfil\break\noindent
{\it  4-}~M.Dutta,S.Mahapatra,D.Borah,N.Sahu, 
\hfil\hfil\break\noindent
hep-ph:2101.06472 and references therein
\hfil\hfil\break\noindent
{\it  5-}~J.Billard,M.Boulay,S.Cebrian,L.Covi, {\it APPEC Committee Report};
hep-ex:2104.07634 and references therein
\hfil\hfil\break\noindent
{\it  6-}~M.Szydagis,C.Levy,G.M.Blockinger,A.Kamaha,N.
\hfil\hfil\break\noindent
Parveen,G.R.C.Rischbieter,
Phys.Rev.{\bf D103}(2021)
\hfil\hfil\break\noindent
012002;hep-ex:2007.00528
\hfil\hfil\break\noindent
{\it  7-}~A.E.Robinson, hep-ex:2006.13278
\hfil\hfil\break\noindent
{\it  8-}~E.Aprile,{\it et als}, XENON-Collaboration, JCAP{\bf 11}
\hfil\hfil\break\noindent
(2020)03;physics.ins-det:2007.080796
\hfil\hfil\break\noindent
{\it  9-}~D.S.Akerib, {\it et als}, LUX-Collaboration,
\hfil\hfil\break\noindent
Phys. Rev.{\bf D101}(2020)052002;
astro-ph.IM:1802.06039
\hfil\hfil\break\noindent
{\it  10-}A.H.Abdelhameed,{\it et als}, CRESST-Collaboration,
\hfil\hfil\break\noindent
Phys.Rev.{\bf D100}(2019)102002;
astro-ph.CO:1904-00498
\hfil\hfil\break\noindent
{\it  11-}~H.Kluck, CRESST-Collaboration, J.Phys.Conf.
\hfil\hfil\break\noindent
Serv.{\bf 1468}(2020)012038
\hfil\hfil\break\noindent
{\it  12-}~X.Ren, {\it et als},PANDA-Collaboration,
\hfil\hfil\break\noindent
Phys.Rev.Lett.{\bf 121}(2018)021304
hep-ex:1802.06912
\hfil\hfil\break\noindent
{\it  13-}~J.Yang, {\it et als}, PANDA-Collaboration; 
\hfil\hfil\break\noindent
hep-ex:2104.14724
\hfil\hfil\break\noindent
{\it  14-}~H.An., M.Pospelov, J.Pradler, A.Ritz, 
\hfil\hfil\break\noindent
Phys.Lett.{\bf B747}(2015)331-338;
hep-ph:1412.8378
\hfil\hfil\break\noindent
{\it  15-}~G.Pr\'ezeau, A.Kurylov, M.Kamionsky, P.Vogel, Phys.Rev.Lett.{\bf 91}(2003),231301;
astro-ph:0309115
\hfil\hfil\break\noindent
{\it  16-}~D. Hooper, S.M. McDermott, Phys.Rev.{\bf D97}
\hfil\hfil\break\noindent
(2018)115006; hep-ph:1802.03025
\hfil\hfil\break\noindent
{\it  17-}~E.Behnke, {\it et als}, PICASSO-Collaboration, 
\hfil\hfil\break\noindent
Astropart.Phys.{\bf 90}(2017)85-92;
hep-ex:1611:01499
\hfil\hfil\break\noindent
{\it  18-}~M.Lafreni\`ere, Th\`ese de Doctorat, D\'ept. de Physique, U.de Montr\'eal, D\'ec. 2016;
and references therein.
\hfil\hfil\break\noindent
{\it  19-}~C.Amole, {\it et als},PICO-Collaboration, 
\hfil\hfil\break\noindent
Phys.Rev.{\bf D100}(2019)022001;
astro-ph.CO;1902.04031
\hfil\hfil\break\noindent
{\it  20-}~C.B. Krauss, PICO-Collaboration, 
\hfil\hfil\break\noindent
J.Phys.Ser.{\bf 1468}(2020)012043
\hfil\hfil\break\noindent
{\it  21-}~R. Bernabei,{\it et als.},Int.J.Mod.Phys.{\bf A22}(2007)3155-3168;astro-ph:0706.1421
\hfil\hfil\break\noindent
{\it  22-}~M.Ibe, W.Nakano, Y.Shoji, K.Suzuki, 
\hfil\hfil\break\noindent
JHEP{\bf 03}(2018),194; hep-ph:1707.07258
\hfil\hfil\break\noindent
{\it  23-}~N.F.Bell,J.B.Dent,J.L.Newstead,S.Sabharwal,T.J.Weiler, Phys.Rev.{\bf D101}(2020)
015012; hep-ph:1905.00046
\hfil\hfil\break\noindent
{\it  24-}~V.S.Flambaum, L.Su, L.Wu, B,Zhu; 
\hfil\hfil\break\noindent
hep-ph:2012.09751
\hfil\hfil\break\noindent
{\it  25-}~J.A.Dror, G.Elor, R.McGehee, T.-T.Yu, 
\hfil\hfil\break\noindent
Phys.Rev.{\bf D103}(2021)035001;
hep-ph:2011.01940
\hfil\hfil\break\noindent
{\it  26-}~C.Kouvaris, J.Pradler, Phys.Rev.Lett.{\bf 118}(2017)
\hfil\hfil\break\noindent
03180; hep-ph:1607.01789
\hfil\hfil\break\noindent
{\it  27-}~M.J.Dolan, F.Kahlhoefer, C.McCabe, 
\hfil\hfil\break\noindent
Phys.Rev.Lett.{\bf 121}(2018),101801;
hep-ph:1711.09906
\hfil\hfil\break\noindent
{\it  28-}~R.Essig, J.Mardon, T.Volansky, JHEP{\bf 05}(2016)046; hep-ph:1108.5383
\hfil\hfil\break\noindent
{\it  29-}~B.V.Lehmann, S. Profumo, Phys.Rev.{\bf D102}(2020)
\hfil\hfil\break\noindent
023938; hep-ph:2002.07809
\hfil\hfil\break\noindent
{\it  30-}~G.D.Starkman, D.N.Spergel, Phys.Rev.Lett.{\bf 74} 
\hfil\hfil\break\noindent
(1995)623-2625; hep-ph:9409275
\hfil\hfil\break\noindent
{\it  31-}~T. Bringmann, M. Pospelov, Phys.Rev.Lett.{\bf 122}
\hfil\hfil\break\noindent
(2019)171801; hep-ph:1810.10543
\hfil\hfil\break\noindent
{\it  32-}~J.Alvey,M.Campos,M.Fairbairn,T.You,
\hfil\hfil\break\noindent
Phys.Rev.Lett.{\bf 123}(2019)261802;
hep-ph:1905.05776
\hfil\hfil\break\noindent
{\it  33-}~M.Klasen,M.Pohl,G.Sigi, Prog.in Part. and Nucl.
\hfil\hfil\break\noindent
Phys.{\bf 85}(2015)1-32;hep-ph:1507.03800
\hfil\hfil\break\noindent
{\it  34-}~N.Vinyoles,A.Serenelli,F.L.Villante,S.Basu,
\hfil\hfil\break\noindent
J.Redondo,J.Isern,JCAP{\bf 10}(2015),015;
\hfil\hfil\break\noindent
astro-ph.SR:1501.01639
\hfil\hfil\break\noindent
{\it  35-}~H.An, M. Pospelov, J.Pradler, A. Ritz, 
\hfil\hfil\break\noindent
Phys.Rev.Lett.{\bf 120}(2018)141801;
hep-ph:1708.03642
\hfil\hfil\break\noindent
{\it  36-}~R.K.Leane, J.Smirnov, Phys.Rev.Lett.{\bf 126}
\hfil\hfil\break\noindent
(2021)161101; hep-ph:2021.00015
\hfil\hfil\break\noindent
{\it  37-}~V.Gluscevic, K.K.Boddy, Phys.Rev.Lett.
{\bf 121}(2018)
\hfil\hfil\break\noindent
011301;astro-ph.CO:1712.07133
\hfil\hfil\break\noindent
{\it  38-}~Y.Ali-Ha\"\i moud, J.Chuba, M.Kamionskowski, 
\hfil\hfil\break\noindent
Phys.Rev.Lett.,{\bf 115}071304;
astro-ph.CO:1506.04745
\hfil\hfil\break\noindent
{\it  39-}~K. Maamari,V.Gluscevic,K.K.Boddy,E.O.Nadler,
\hfil\hfil\break\noindent
R.H.Wechsler,Astrophys.J.Lett.,
{\bf 907}(2021)2,L46;
\hfil\hfil\break\noindent
astro-ph.CO:2010.02936
\hfil\hfil\break\noindent
{\it  40-}~S.Afach, {\it et als.},astro-ph.CO:210213379
\hfil\hfil\break\noindent
{\it  41-}~A.Derevianko, M. Pospelov, Nature,{\bf Physics 10}
\hfil\hfil\break\noindent
(2014)933;
physics.atom-ph:1311:1244
\hfil\hfil\break\noindent
{\it  42-}~M.Bauer, M.Heiles, M.Neubert, A.Thamm, 
\hfil\hfil\break\noindent
Eur.Phys.J.{\bf C79}(2019)74;
\hfil\hfil\break\noindent
{\it  43-}~H.Mies,C.Schreb,P.Schwaller,JHEP{\bf 04}(2021)049;hep-ph:2100.1399
\hfil\hfil\break\noindent
{\it  44-}~J.I.Read, J.Phys.{\bf G41}(2014)063101;astro-ph.GA:
\hfil\hfil\break\noindent
1404.1938
\hfil\hfil\break\noindent
{\it  45-}~C.J.Copi,L.M.Krauss,Phys.Rev.{\bf D63}(2001),043507; 
\hfil\hfil\break\noindent
astro-ph:0009467
\hfil\hfil\break\noindent
{\it  46-}~P.Salucci, F.Nesti, G.Gentile, C.F.Martins, 
\hfil\hfil\break\noindent
Astron.Astrophysics,{\bf 523}(2010)A83
\hfil\hfil\break\noindent
{\it  47-}~H.Bethe, W.Heitler, Proc.Roy.Soc.{\bf 146}(1934)83
\hfil\hfil\break\noindent
{\it  48-}{\it Classical Electrodynamics}, J.D. Jackson; Wiley, New York, $2^{nd}$ Edition, 1975
\hfil\hfil\break\noindent
{\it  49-}~P.Talukdar, F.Myhrer, U.Raha,Eur.Phys.J.{\bf A54}
\hfil\hfil\break\noindent
(2018)11,195; nucl-th:1712.09963
\hfil\hfil\break\noindent
{\it  50-}~U.Eichmann,W.Greiner, J.Phys.{\bf G23}(1997)L65-L76; nucl-th:9706044
\hfil\hfil\break\noindent
{\it  51-}~T.S.Biro,K.Nita,A.L.DePaoli,W.Bauer,W.Cassing,
\hfil\hfil\break\noindent
U.Mosel,Nucl.Phys.{\bf A475}(1987)579-597
\hfil\hfil\break\noindent
{\it  52-}~G.Baur, A.Leuschner. Eur.Phys.J.{\bf C8}(1999)
\hfil\hfil\break\noindent
631-635;hep-ph:9902245
\hfil\hfil\break\noindent
{\it  53-}~W. Zhu, Nucl.Phys.`{\bf B953}(2020)114958, 
\hfil\hfil\break\noindent
hep-ph:1909.03053
\hfil\hfil\break\noindent
{\it  54-}~M.Cirelli, P.D.Serpico, G.Zaharijas, JCAP{\bf 11}(2013)
\hfil\hfil\break\noindent
035; astro-ph.HE:1307.7152
\hfil\hfil\break\noindent
{\it  55-}~J.B.Dent,B.Dutta,J.L.Newstead,A.Rodriguez,I.M.
\hfil\hfil\break\noindent
Shoemaker,Z.Tabrizi; hep-ph:2012,07930
\hfil\hfil\break\noindent
{\it  56-}~L.Su, L.Wu, and B.Zhu; hep-ph:2105-06326
\hfil\hfil\break\noindent
{\it  57-}~{\it Relativistic Quantum Mechanics}, Bjorken J.D. and S.D. Drell; Vol. 1, McGraw-Hill,
New York, 1964
\hfil\hfil\break\noindent
{\it  58-}~{\it An Introduction to Quantum Field Theory}, Peskin, M.E. and D. V. Schroeder;
Addison Wesley, New York, 1995
\hfil\hfil\break\noindent
{\it  59-}~{\it Relativistic Quantum Mechanics and Field Theory}, Gross, F.,Wiley Interscience, 
New York, 1993
\hfil\hfil\break\noindent
{\it  60-}~{\it Relativistic Kinematics}, R. Hagedorn; 
\hfil\hfil\break\noindent
Benjamin/Cummings, London, 1963
\hfil\hfil\break\noindent
{\it  61-}~ {\it Collider Physics}, Barger, V.D. and R.J.N. Phillips, Addison Wesley, 
New York, 1987
\hfil\hfil\break\noindent
{\it  62-}~A.Savitzky, M.J.E.Golay, Analyt.Chem.{\bf 8}(1964)1627-1639
\hfil\hfil\break\noindent
{\it  63-}~W.Herr, B.Muratori,{\it CERN Acc. School and DESY Zeuthen: Accelerator Physics},
(2003)361-377
\hfil\hfil\break\noindent
{\it  64-}~H.Burkhardt, P. Grafstrom, LHC Project Report 1019 (2007)
\hfil\hfil\break\noindent
{\it  65-}~M. Pospelov, A. Ritz, M. Voloshin, Phys.Lett.{\bf B662} (2008)53-61;
hep-ph:0711.4866
\hfil\hfil\break\noindent
{\it  66-}~K.Schutz,T.R.Statyer, JCAP{\bf 01}(2015),21;
\hfil\hfil\break\noindent
hep-ph:1409.2867
\hfil\hfil\break\noindent
{\it  67-}~ A.Dedes, I.Giomataris, K. Suxho, J.D.Vergados,
\hfil\hfil\break\noindent
Nucl.Phys.{\bf B826} (2010)148-177;
hep-ph:0907.0758
\hfil\hfil\break\noindent
{\it  68-}~T.R.Lauer, {\it et als}, Astrophys.J.{\bf 96}(2021)2,77;
\hfil\hfil\break\noindent
astro-ph.GA:2011.03052
\hfil\hfil\break\noindent
{\it  69-}~N.F.Bell,Y.Cai,J.H.Dent,R.K.Leane,T.Weiler, 
\hfil\hfil\break\noindent
Phys.Rev.{\bf D96}(2017),023011;
hep-ph:1705.01105
\hfil\hfil\break\noindent
{\it  70-}~R. Bernabei, {\it et als.}, Phys.Rev.{\bf D77} (2008)
\hfil\hfil\break\noindent
023506;hep-ph:0712.0562
\hfil\hfil\break\noindent
{\it  71-}~J.F.Beacom, N.F. Bell, G.Bertone, Phys.Rev.Lett.
\hfil\hfil\break\noindent
{\bf 94}(2005)171301; astro-ph:0409403
\hfil\hfil\break\noindent
{\it  72-}~A.Coogan,A.Moiseev,L.Morrison,S.Profumo, 
\hfil\hfil\break\noindent
astro-ph.HE:2101.10370
\hfil\hfil\break\noindent
{\it  73-}~D.Gaggero, M.Valli, Adv.in High Energy Phys.
\hfil\hfil\break\noindent
{\bf 2018}(2018)3010514;astro-ph.HE:1802.00636
\hfil\hfil\break\noindent
\hfil\hfil\break
\hfil\hfil\break
\noindent{\bf{Table 1}}\hfil\hfil
\hfil\hfil\break\noindent
Integrated expected photon fluxes for the two scenarios considered, for
different observation angles, and for different energy bands.
In scenario A, the colliding particle has a
mass of 2 GeV while the exchange particle has a mass of 0.1 GeV, while
both have a mass of 0.1 GeV in scerario B. A-180-01 means: scenario A,
observation angle of 180 degress (directly at the sun) and observation point
at 1 au from the sun. The fluxes are in $1/m^2 sec$. If one takes into
account that the opening angle of the observation cone is $40^\circ$, then
our units for the flux are $1/m^2 sec (0.47~\pi~sr)$
\vfil\vfil\eject
\begin{figure}
    \centering
    \includegraphics[scale=0.5]{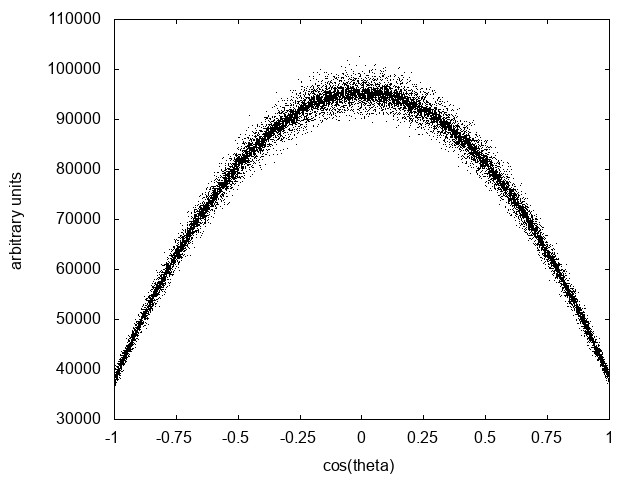}
    \caption{Angular distribution before and after smoothing}
    \label{figure 1}
\end{figure}
\begin{figure}
    \centering
    \includegraphics[scale=0.5]{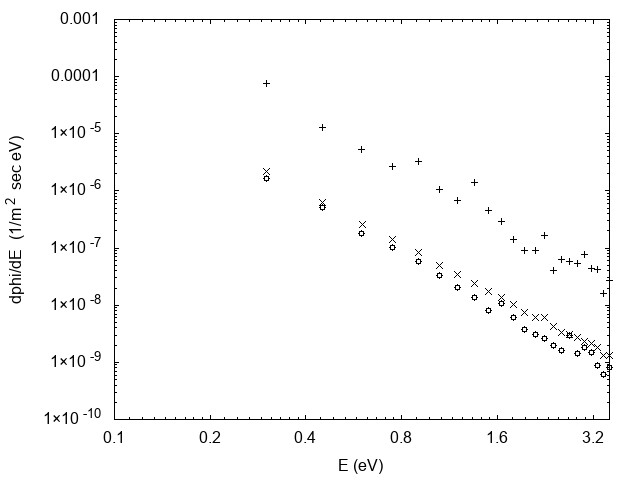}
    \caption{Photon flux ($1/m^2~sec~eV$) {\it vs} photon energy (eV) for 1au-180 degrees (+), 1au-45 degrees (x), and 50au-180 degrees (o) in scenario A}
    \label{figure 2}
\end{figure}
\begin{figure}
    \centering
    \includegraphics[scale=0.5]{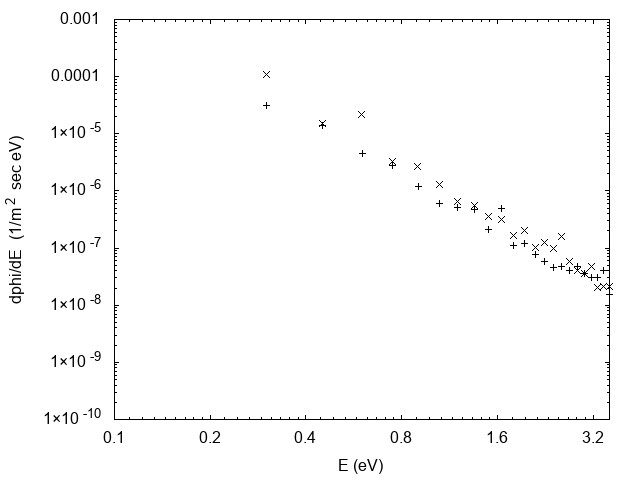}
    \caption{Photon flux ($1/m^2~sec~eV$) {\it vs} photon energy (eV) for 1au-90 degrees (+), and 1au-135 (x) in scenario A}
    \label{figure 3}
\end{figure}
\begin{figure}
    \centering
    \includegraphics[scale=0.5]{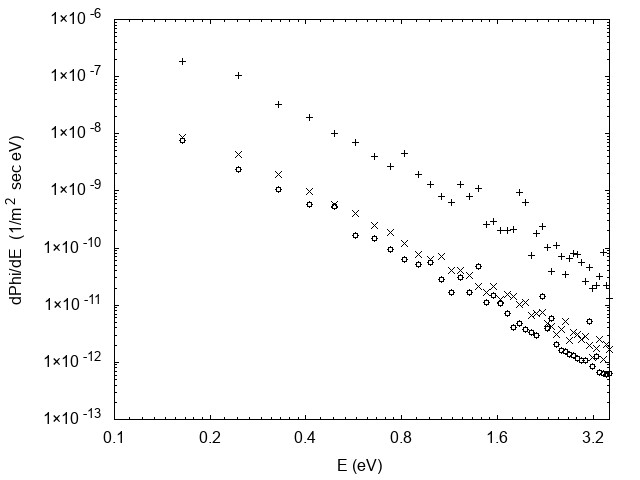}
    \caption{Photon flux ($1/m^2~sec~eV$) {\it vs} photon energy (eV) for 1au-180 degrees (+), 1au-45 degrees (x) and 50au-180 degrees (o) in scenario B}
    \label{figure 4}
\end{figure}
\begin{figure}
    \centering
    \includegraphics[scale=0.5]{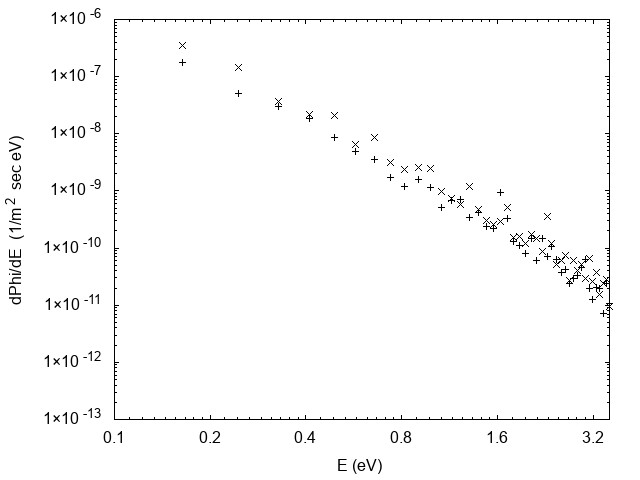}
    \caption{Photon flux ($1/m^2~sec~eV$) {\it vs} photon energy (eV) for 1au-90 degrees (+), and 1au-135 degrees (x) in scenario B}
    \label{figure 5}
\end{figure}
\vfil\vfil\eject
{\hsize=7.0truein
$$
\vbox{\offinterlineskip
\halign{&\vrule#&\strut\ #\ \cr
\multispan{17}\hfil\bf Table I \hfil\cr
\hfil\cr
\noalign{\smallskip}
\noalign{\hrule}
height3pt&\omit&&\omit&&\omit&&\omit&&\omit&&\omit&&\omit&&\omit&\cr
&\hfil scenario\hfil&&\hfil slope\hfil&&\hfil
1.6-3.2 eV\hfil&&\hfil 0.5-1.0 eV\hfil&&\hfil
 0.1-0.5 eV\hfil&&\hfil 10-50 meV\hfil&&
\hfil 1-5 meV\hfil&&\hfil 0.1-0.5 meV\hfil&\cr
height3pt&\omit&&\omit&&\omit&&\omit&&\omit&&\omit&&\omit&&\omit&\cr
\noalign{\hrule}
\noalign{\hrule}
height3pt&\omit&&\omit&&\omit&&\omit&&\omit&&\omit&&\omit&&\omit&\cr
&\hfil A-180-01\hfil&&\hfil -3.38\hfil&&\hfil
$9.5\times 10^{-8}$\hfil&&\hfil $1.5\times 10^{-6}$\hfil&&\hfil
 $8.4\times 10^{-5}$\hfil&&\hfil $0.021$\hfil&&
\hfil $5.1$\hfil&&\hfil $1200$\hfil&\cr
height3pt&\omit&&\omit&&\omit&&\omit&&\omit&&\omit&&\omit&&\omit&\cr
\noalign{\hrule}
height3pt&\omit&&\omit&&\omit&&\omit&&\omit&&\omit&&\omit&&\omit&\cr
&\hfil A-135-01\hfil&&\hfil -3.50\hfil&&\hfil
$1.4\times 10^{-7}$\hfil&&\hfil $2.6\times 10^{-6}$\hfil&&\hfil
 $1.7\times 10^{-4}$\hfil&&\hfil $0.053$\hfil&&
\hfil $17$\hfil&&\hfil $5400$\hfil&\cr
height3pt&\omit&&\omit&&\omit&&\omit&&\omit&&\omit&&\omit&&\omit&\cr
\noalign{\hrule}
height3pt&\omit&&\omit&&\omit&&\omit&&\omit&&\omit&&\omit&&\omit&\cr
&\hfil A-90-01\hfil&&\hfil -3.10\hfil&&\hfil
$1.0\times 10^{-7}$\hfil&&\hfil $1.14\times 10^{-6}$\hfil&&\hfil
 $4.2\times 10^{-5}$\hfil&&\hfil $5.4\times 10^{-3}$\hfil&&
\hfil $0.68$\hfil&&\hfil $86$\hfil&\cr
height3pt&\omit&&\omit&&\omit&&\omit&&\omit&&\omit&&\omit&&\omit&\cr
\noalign{\hrule}
height3pt&\omit&&\omit&&\omit&&\omit&&\omit&&\omit&&\omit&&\omit&\cr
&\hfil A-45-01\hfil&&\hfil -2.99\hfil&&\hfil
$8.4\times 10^{-9}$\hfil&&\hfil $8.6\times 10^{-8}$\hfil&&\hfil
 $2.7\times 10^{-6}$\hfil&&\hfil $2.6\times 10^{-4}$\hfil&&
\hfil $0.026$\hfil&&\hfil $2.6$\hfil&\cr
height3pt&\omit&&\omit&&\omit&&\omit&&\omit&&\omit&&\omit&&\omit&\cr
\noalign{\hrule}
height3pt&\omit&&\omit&&\omit&&\omit&&\omit&&\omit&&\omit&&\omit&\cr
&\hfil A-180-50\hfil&&\hfil -3.28\hfil&&\hfil
$4.5\times 10^{-9}$\hfil&&\hfil $6.3\times 10^{-8}$\hfil&&\hfil
 $3.0\times 10^{-6}$\hfil&&\hfil $5.9\times 10^{-4}$\hfil&&
\hfil $0.11$\hfil&&\hfil $22$\hfil&\cr
height3pt&\omit&&\omit&&\omit&&\omit&&\omit&&\omit&&\omit&&\omit&\cr
\noalign{\hrule}
\noalign{\hrule}
height3pt&\omit&&\omit&&\omit&&\omit&&\omit&&\omit&&\omit&&\omit&\cr
&\hfil B-180-01\hfil&&\hfil -3.17\hfil&&\hfil
$1.2\times 10^{-10}$\hfil&&\hfil $1.5\times 10^{-9}$\hfil&&\hfil
 $6.2\times 10^{-8}$\hfil&&\hfil $9.1\times 10^{-6}$\hfil&&
\hfil $1.4\times 10^{-3}$\hfil&&\hfil $0.20$\hfil&\cr
height3pt&\omit&&\omit&&\omit&&\omit&&\omit&&\omit&&\omit&&\omit&\cr
\noalign{\hrule}
height3pt&\omit&&\omit&&\omit&&\omit&&\omit&&\omit&&\omit&&\omit&\cr
&\hfil B-135-01\hfil&&\hfil -3.16\hfil&&\hfil
$1.5\times 10^{-10}$\hfil&&\hfil $1.8\times 10^{-9}$\hfil&&\hfil
 $7.2\times 10^{-8}$\hfil&&\hfil $1.0\times 10^{-5}$\hfil&&
\hfil $1.5\times 10^{-3}$\hfil&&\hfil $0.21$\hfil&\cr
height3pt&\omit&&\omit&&\omit&&\omit&&\omit&&\omit&&\omit&&\omit&\cr
\noalign{\hrule}
height3pt&\omit&&\omit&&\omit&&\omit&&\omit&&\omit&&\omit&&\omit&\cr
&\hfil B-90-01\hfil&&\hfil -3.10\hfil&&\hfil
$8.5\times 10^{-11}$\hfil&&\hfil $9.7\times 10^{-10}$\hfil&&\hfil
 $3.6\times 10^{-8}$\hfil&&\hfil $4.6\times 10^{-6}$\hfil&&
\hfil $5.7\times 10^{-4}$\hfil&&\hfil $0.072$\hfil&\cr
height3pt&\omit&&\omit&&\omit&&\omit&&\omit&&\omit&&\omit&&\omit&\cr
\noalign{\hrule}
height3pt&\omit&&\omit&&\omit&&\omit&&\omit&&\omit&&\omit&&\omit&\cr
&\hfil B-45-01\hfil&&\hfil -2.96\hfil&&\hfil
$8.9\times 10^{-12}$\hfil&&\hfil $8.7\times 10^{-11}$\hfil&&\hfil
 $2.6\times 10^{-9}$\hfil&&\hfil $2.4\times 10^{-7}$\hfil&&
\hfil $2.2\times 10^{-5}$\hfil&&\hfil $0.002$\hfil&\cr
height3pt&\omit&&\omit&&\omit&&\omit&&\omit&&\omit&&\omit&&\omit&\cr
\noalign{\hrule}
height3pt&\omit&&\omit&&\omit&&\omit&&\omit&&\omit&&\omit&&\omit&\cr
&\hfil B-180-50\hfil&&\hfil -3.13\hfil&&\hfil
$3.4\times 10^{-12}$\hfil&&\hfil $4.1\times 10^{-11}$\hfil&&\hfil
 $1.6\times 10^{-9}$\hfil&&\hfil $2.1\times 10^{-7}$\hfil&&
\hfil $2.9\times 10^{-5}$\hfil&&\hfil $0.004$\hfil&\cr
height3pt&\omit&&\omit&&\omit&&\omit&&\omit&&\omit&&\omit&&\omit&\cr
\noalign{\hrule}
\noalign{\hrule}\noalign{\smallskip}}}
$$}
\end{document}